\documentclass[aip, twocolumn,  showpacs, superscriptaddress, 10pt]{revtex4-1}

\usepackage{amsmath,amssymb,graphicx,natbib}
\usepackage{graphicx}% Include figure files
\usepackage{dcolumn}% Align table columns on decimal point
\usepackage{bm}% bold math
\usepackage{epstopdf}
\begin{document}

%\preprint{AIP/123-QED}

\title{A nanowaveguide platform for collective atom-light interaction}

\author{Y. Meng}
\altaffiliation{Y. Meng and J. Lee contributed equally to this work.}
\affiliation{Department of Electrical and Computer Engineering, University of Maryland, College Park, Maryland 20742, USA}

\author{J. Lee}
\altaffiliation{Y. Meng and J. Lee contributed equally to this work.}
\affiliation{Joint Quantum Institute, Department of Physics, University of Maryland and NIST, College Park, Maryland 20742, USA}

\author{M. Dagenais}
\email{dage@ece.umd.edu.}
\affiliation{Department of Electrical and Computer Engineering, University of Maryland, College Park, Maryland 20742, USA}

\author{S. L. Rolston}
\affiliation{Joint Quantum Institute, Department of Physics, University of Maryland and NIST, College Park, Maryland 20742, USA}

%\date{\today}

\begin{abstract}
We propose a nanowaveguide platform for collective atom-light interaction through  evanescent field coupling. We have developed a 1\,cm-long silicon nitride nanowaveguide can use evanescent fields to trap and probe an ensemble of $^{87}$Rb atoms. The waveguide has a sub-micrometer square mode area and was designed with tapers for high fiber-to-waveguide coupling efficiencies at near-infrared wavelengths (750\,nm to 1100\,nm). Inverse tapers in the platform adiabatically transfer a weakly guided mode of fiber-coupled light into a strongly guided mode with an evanescent field to trap atoms and then back to a weakly guided mode at the other end of the waveguide. The coupling loss is $-1$\,dB per facet ($\sim$80\,\% coupling efficiency) at 760\,nm and 1064\,nm, which is estimated by a propagation loss measurement with waveguides of different lengths. The proposed platform has good thermal conductance and can guide high optical powers for trapping atoms in ultra-high vacuum. As an intermediate step, we have observed thermal atom absorption of the evanescent component of a nanowaveguide, and have demonstrated the U-wire mirror magneto-optical trap that can transfer atoms to the proximity of the surface.
\end{abstract}

%\pacs{Valid PACS appear here}
\maketitle

Atom-light interactions can be harnessed for a number of quantum-based applications,  such as quantum information processing~\cite{Porto03, Lett09, Tittle09, Jessen99, Kimble08, Choi08} and atomic sensing~\cite{Kasevich13, Biedermann11, Romalis03}. Enhancing the atom-light interaction with small mode areas has been demonstrated in several platforms such as  hollow-core fibers~\cite{Gaeta10, Gaeta11},  hollow-core waveguides~\cite{Yang07, Wu10},  tapered fibers~\cite{Spillane08, Hendrickson10} and a nanowaveguide~\cite{Levy13}. An array of a few micrometer-square free-space optical spots in a trench from an optical waveguide array were also recently used to interrogate ultracold atoms formed by a mirror magneto-optical trap (MOT) above the trench~\cite{Hinds11}. In contrast to thermal alkali vapors~\cite{Gaeta10, Gaeta11, Yang07, Wu10, Spillane08, Hendrickson10, Levy13} or a free-space mode from an optical  waveguide array~\cite{Hinds11}, laser-cooled ultracold atoms trapped in the evanescent field of a waveguide can exhibit much stronger coupling between the atoms and photons due to good overlap of the atoms with the small optical mode, and the lack of Doppler broadening.  In addition, cold trapped atoms have a longer residence time in the field and thus a longer  coherence time. All-optical switching with optically trapped atoms has previously been implemented in a hollow fiber~\cite{Lukin09}, and nanofiber-based optical lattices for $^{133}$Cs and $^{87}$Rb atoms were also demonstrated~\cite{Vetsch10, Lee15}. Although nanofiber atom traps create strong atom-light interactions, they are not easily scalable to realize  complex photonic circuits, while an integrated optics approach with waveguides enjoys both strong interactions and the potential for scalability.

Here we present a potentially scalable nanowaveguide platform, which can operate as the atom-light interface between an evanescent field probe and cold neutral atoms. An open-window in the middle of the waveguide is used to trap and probe atoms through the evanescent field. Atoms loaded from an atom-chip mirror MOT  will be optically-trapped with a two-color evanescent field atom trap provided by red-detuned counterpropagating waveguide fields and a blue-detuned traveling waveguide field with van der Waals potential. For the atom trap, it is necessary to perform measurements characterizing the optical properties of such waveguides, focusing on designs that can simultaneously support the two optical frequencies.

\begin{figure}
\centering\includegraphics[width=1\columnwidth]{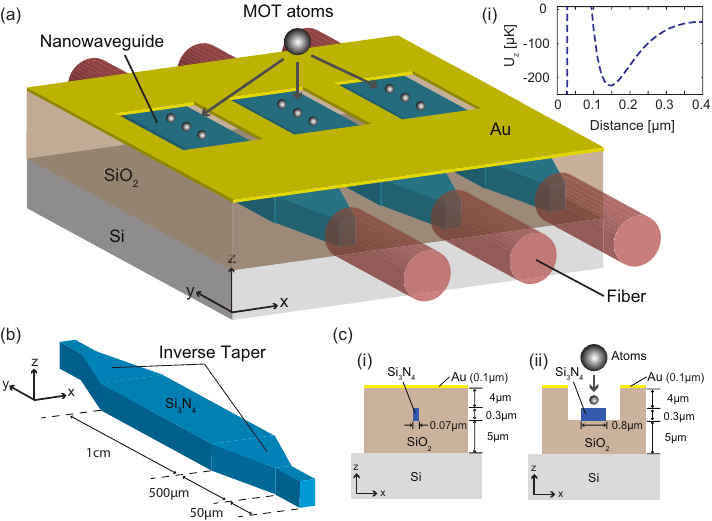}
\caption{The concept of a nanowaveguide platform for collective atom-light interaction. (a) Integration of a nanowaveguide and a mirror MOT (3D-plot, not-to-scale) (i) The total potential along the vertical direction of the waveguide for Rb atoms ($\lambda_{\mathrm{Rb}}$ = 780\,nm) is $U_{tot} = \frac{3\pi c^2}{2\omega_{0}^3}\left( \frac{\Gamma}{\Delta_{blue}} I_{blue}(\mathbf{r}) + \frac{\Gamma}{\Delta_{red}} I_{red}(\mathbf{r})\right) - \frac{C_{vdW}}{z^3}$, where optical powers of a blue-detuned traveling waveguide field (760\,nm) and one of red-detuned counterpropagating fields (1064\,nm) are $P_{blue}\,\simeq$\,12\,mW and $P_{red}\,\simeq$\,10\,mW. (b) Silicon nitride ($\mathrm{Si_{3}N_{4}}$) core and inverse tapers of the nanowaveguide (3D-plot, not-to-scale). (c) 2-D geometry of the nanowaveguide (i) Side view of the nanowaveguide input (xz plane, not-to-scale). (ii) Cross-sectional view of the nanowaveguide at the open-window (xz plane, not-to-scale); the width of the open-window is $\simeq$\,20\,$\mu$m.}
\label{fig_concept}
\end{figure}

Fig.~\ref{fig_concept} (a) shows the schematic image of a nanowaveguide platform, and Fig.~\ref{fig_fab_prop_loss} (a), (b) shows the image of the real waveguide sample. Instead of using a nanofiber~\cite{Vetsch10, Lee15} or a nanophotonic crystal waveguide~\cite{Lukin13, Kimble14}, we consider an integrated design of a nanowaveguide and an atom chip mirror MOT~\cite{Reichel99} which has the advantage of better heat dissipation and good scalability. The transporting and loading procedure from a mirror MOT above the surface to the free-space mode of an optical waveguide having a few micrometer square area ($4\times4\,\mu m^{2}$) in the trench structure was previously demonstrated~\cite{Hinds11}. In our experiment, nanowaveguide atom trapping with a sub-micrometer square evanescent field mode will be possible and the two-color evanescent field atom trap minimum will be around 150\,nm above the waveguide surface~\cite{Rolston13} (see Fig.~\ref{fig_concept} (a)(i)).

\begin{figure}
\centering\includegraphics[width=1\columnwidth]{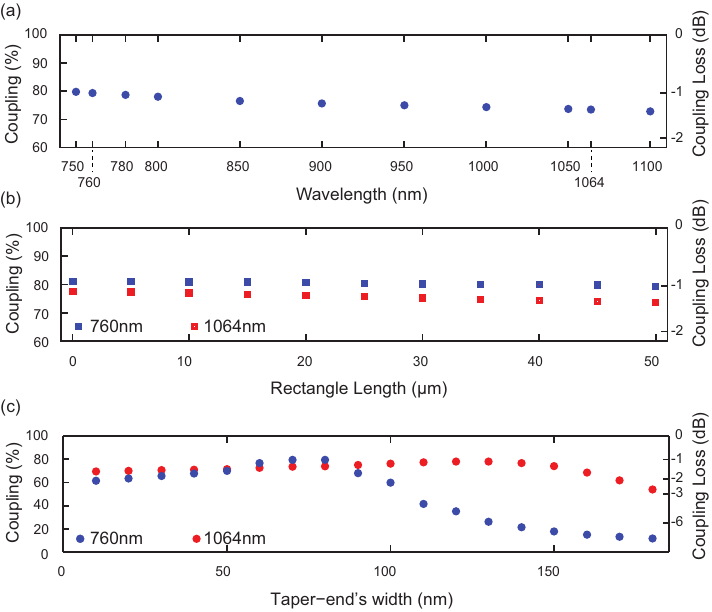}
\caption{The simulation of the fiber-to-waveguide coupling with the TE mode (a) The wavelength vs. the coupling efficiency with 500\,$\mu$m long and 70\,nm wide tapers. (b) The rectangle length at the end of inverse tapers vs. the coupling efficiency for 760\,nm (blue) and 1064\,nm (red). (c) The width of inverse taper-ends vs. the coupling efficiency for 760\,nm (blue) and 1064\,nm (red) with an extra 50\,$\mu$m-long rectangle.}
\label{fig_design_sim}
\end{figure}

\begin{figure}
\centering\includegraphics[width=1\columnwidth]{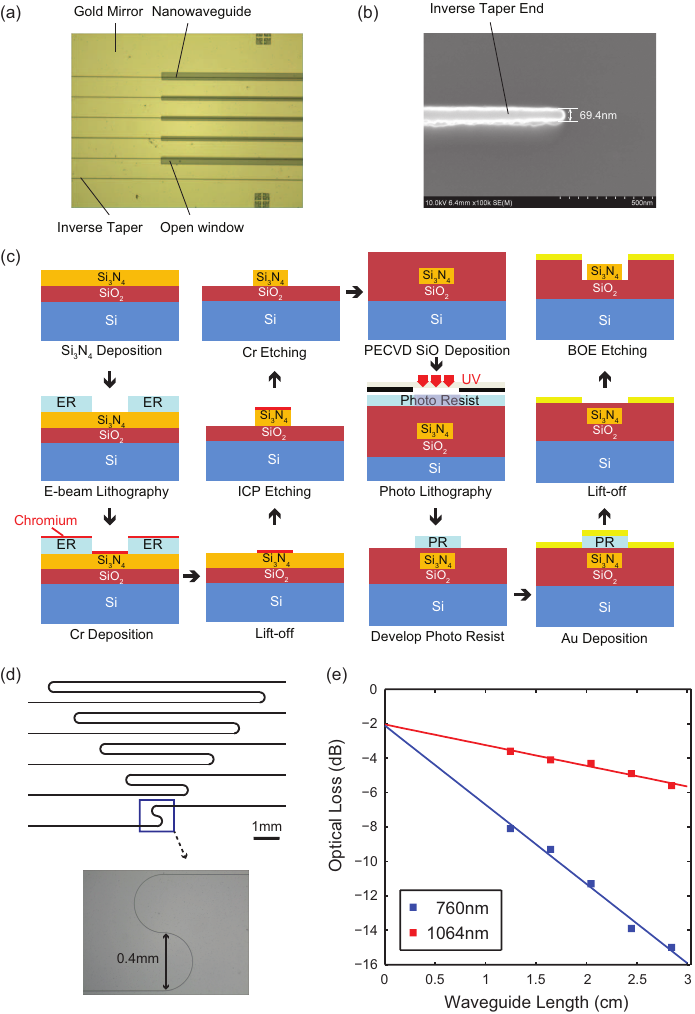}
\caption{Nanowaveguide fabrication and propagation loss measurement (a) The real waveguide sample; taper length 500 $\mu$m, waveguide length = 1cm. (b) The SEM image of an inverse taper-end. (c) Fabrication Process. (d) Propagation loss measurement patterns with multiple waveguides in different lengths; inset is the image of a waveguide pattern with a bending. (e) The optical loss  vs. the waveguide length. The estimated propagation losses from fitting lines are $-4.6$\,dB/cm at 760\,nm and $-1.2$\,dB/cm at 1064\,nm. The total coupling losses from the y-axis intersection points are $-2.11$\,dB at 760\,nm and $-2.044$\,dB at 1064\,nm. The coupling losses are $-1.055$\,dB per facet (78.4\,\% coupling efficiency) at 760\,nm and $-1.022$\,dB per facet (79\,\% coupling efficiency) at 1064\,nm.}
\label{fig_fab_prop_loss}
\end{figure}

Using the simulation software, FimmProp (see Fig.~\ref{fig_design_sim}), we have designed a nanowaveguide (800\,nm$\times$300\,nm, $n_{\mathrm{Si_{3}N_{4}}}$ = 2 and $n_{\mathrm{SiO_2}}$ = 1.46) that guides 760\,nm blue-detuned trapping light, 1064\,nm red-detuned trapping light, and 780\,nm probe light for $^{87}$Rb atoms~\cite{Rolston13}. An inverse-tapered structure that expands the waveguide mode is widely used as a mode converter, and can offer good mode matching and coupling efficiency between a nanowaveguide and an optical fiber. For instance, some previous work on the use of tapers have demonstrated good performance at telecom wavelengths~\cite{Lipson03}, but its advantages and applicability have seldom been reported at near-infrared wavelengths in situations where many wavelengths are used between 750\,nm and 1100\,nm, which is the typical situation used in atomic physics (our waveguide design is also compatible with trapping Cs atoms as well) and in nonlinear optics. Fig.~\ref{fig_design_sim} (a) shows the simulation, which demonstrates that our taper structure can achieve high coupling efficiency for a wide near-infrared wavelength range.

According to the simulation, the horizontal and vertical TE mode field sizes ($1/e^2$ in power) of the waveguide without inverse tapers are (703\,nm$,\,$377\,nm) at 760\,nm and (765\,nm$,\,$474\,nm) at 1064\,nm. For a vacuum compatible platform, we used a fiber-to-waveguide coupling approach based on gluing the fiber to the waveguide chip with UV-epoxy (Epotek OG 116-31). We used Fibercore SM750 (core diameter = 3.82\,$\mu$m, cladding diameter = 125\,$\mu$m, NA = 0.14) fiber, with mode-field diameters of 4.32\,$\mu$m at 760\,nm ($n_{core}$ = 1.46077, $n_{clad}$ = 1.45405) and 5.4\,$\mu$m at 1064\,nm ($n_{core}$ = 1.45635, $n_{clad}$ = 1.44963). The mode mismatch between the fiber and the waveguide would lead to a coupling efficiency of only around 3\,\%, which is too low to provide sufficient intensity to trap atoms with reasonable laser powers. We thus designed an inverse taper to improve the mode matching and increased the measured coupling efficiency. Simulations show that the coupling efficiencies of $\sim$80\,\% and the coupling losses of $-1$\,dB for 760\,nm and 1064\,nm near-infrared wavelengths will be achieved with a taper-end width of 70\,nm and a 500\,$\mu$m taper length that satisfies the adiabaticity condition; Fig.~\ref{fig_design_sim} (c) shows the simulation. A longer taper, such as 1\,mm in length, can increase the coupling efficiency by a few percent, but a longer linear taper also induces more propagation loss and more stitching errors during electron beam lithography (EBL) in practice. Therefore, we chose to work with a $500\,\mu\mathrm{m}$ taper. In addition, the coupling efficiency is very sensitive to the taper-end width especially for 760\,nm as shown in Fig.~\ref{fig_design_sim} (c), which makes cleaving another important issue. Our cleaving technique offers a near perfect mirror-like facet but may have a $\pm20\,\mu\mathrm{m}$ error in cleaving position. If we cleave $40\,\mu\mathrm{m}$ inside the designed taper, the coupling efficiency for 760\,nm will drop to around 25\,\% from 79\,\%. We thus added a rectangular structure at the taper-end to compensate for a potential cleaving error. As we simulated in Fig. 2(b), an additional $50\,\mu\mathrm{m}$-long rectangle only decreases the coupling efficiency by about 3\,\%. In this way, we can reliably achieve a high coupling efficiency.
 
%For a vacuum compatible platform, we uses the fiber-to-waveguide coupling, using UV-epoxy (Epotek OGG 166-31). We use a fiber (Fibercore SM750) with $D_{core}$ = 3.82\,$\mu$m, $D_{clad}$ = 125\,$\mu$m,NA = 0.14, and the $1/e^2$ diameters are 6.34\,$\mu$m at 1064\,nm ($n_{core}$ = 1.45635, $n_{clad}$ = 1.44963) and 4.93\,$\mu$m at 760\,nm ($n_{core}$ = 1.46077, $n_{clad}$ = 1.45405).

The general fabrication process of our nanowaveguide is described in Fig.~\ref{fig_fab_prop_loss} (c). A one-centimeter long $\mathrm{Si_{3}N_{4}}$ waveguide with an inverse taper at both ends is fabricated on a 5\,$\mu$m-thick thermal $\mathrm{SiO_{2}}$ layer (Fig.~\ref{fig_concept} (b)). The 300\,nm $\mathrm{Si_{3}N_{4}}$ was deposited by LPCVD, and EBL was used to pattern the nanowaveguide and create an inverse-tapered structure. Due to our limited EBL writing-field size of order one hundred micrometers, a one-centimeter long waveguide will cross 100 writing fields (each field being 100\,$\mu$m by 100\,$\mu$m) leading to random stitching errors at the writing field boundaries. In order to get a continuous, long waveguide, a Fixed Beam Moving Stage (FBMS) and writing-field overlapping technique were used to minimize  stitching errors. Since we have a 5-$\mu$m-thick $\mathrm{SiO_{2}}$ insulating substrate, another conducting polymer was deposited on top of the ebeam resist (PMMA) to reduce  charging effects. After  e-beam lithography, inductively coupled plasma (ICP) fluorine etching was used to etch the $\mathrm{Si_{3}N_{4}}$. We used PECVD to deposit another 4\,$\mu$m-thick $\mathrm{SiO_{2}}$ above the $\mathrm{Si_{3}N_{4}}$ core. Then with a photolithography pattern and gold deposition, an open-window is made by buffered oxide etch (BOE) in the middle of the sample on the waveguide. The window opening in the center of the waveguide exposes the waveguides to vacuum and thus atoms for trapping. Finally, the sample is cleaved for fiber coupling. Since the atom trap needs to operate in a high vacuum environment, the optical alignment between the fiber and the waveguide should be permanently fixed. Consequently, we glued the fiber onto the waveguide facet with UV epoxy.

To evaluate the coupling efficiency, we  measured the propagation loss of the waveguide. The cut-back technique has been widely used for waveguide propagation loss measurements, but the accuracy of this measurement is highly dependent on the cleave quality. Small differences in the cleaving may lead to large changes in the coupling loss, which then make the propagation loss measurement unreliable. Here we use another method to measure the propagation loss without cleaving the chip more than twice. Figure~\ref{fig_fab_prop_loss} (d) shows the pattern used for this measurement. Each waveguide contains 2 half-circle bends which are exactly the same. They have a 400\,$\mu$m diameter, which is used to minimize the bending loss. Each waveguide is 4\,mm different in length with the neighboring waveguide. After cleaving just once, we can assume that the coupling loss for each waveguide is the same so it offers better accuracy than the traditional cut-back technique. We estimated the coupling efficiencies at 760\,nm and 1064\,nm by the propagation loss measurement with waveguides of different lengths. Assuming both facets have the same coupling efficiency, the coupling losses per facet are $-1$\,dB and the coupling efficiencies per facet are $\sim$80\,\% for 760\,nm and 1064\,nm as shown in Figure~\ref{fig_fab_prop_loss} (e). The measured coupling efficiencies are well matched with the simulation results shown in Fig.~\ref{fig_design_sim}.

\begin{figure}
\centering\includegraphics[width=1\columnwidth]{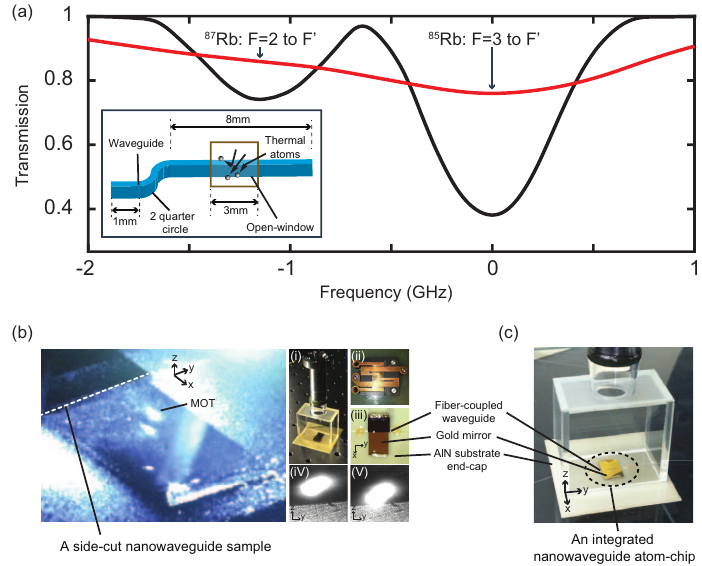}
\caption{(a) Absorption profile of thermal atoms with a L-shaped nanowaveguide. The profile shows $\mathrm{^{85}Rb}$ and $\mathrm{^{87}Rb}$'s D2 transitions. The data (red line) of thermal atoms was measured at $\langle T \rangle \simeq 90^{\circ}\mathrm{C}$ with a nanowaveguide probe ($P \simeq 2\,\mathrm{\mu W}$). The measured data has the total broadening of $2\pi \cdot 765\,\mathrm{MHz}$ (HFHM). The black line is the calculated absorption profile only with Doppler-broadening in a similar way of a reference~\cite{Hughes08}, and the saturation power is $P_{sat} \simeq 25\,\mathrm{nW}$, where $I/I_{sat} = P/P_{sat} \simeq 80$;  The interaction length of the L-shaped waveguide is $l_{int}$ = 3\,mm; the energy of a 780\,nm waveguide probe coupled to thermal atoms is $3.2\,\%$. (Inset) Nanowaveguides with L-shaped geometry and open-windows for thermal atom detection. (b) The U-wire mirror MOT atoms (U-MOT) in a compact UHV glass chamber ($2 \times 4 \times 4\,\mathrm{cm^3}$, $\mathrm{P} \simeq\,1\times10^{-9}\mathrm{mbar}$); the fiber-coupled nanowaveguide (white dotted line) is positioned at the edge of the side-cut nanowaveguide sample. (i) A compact glass chamber with a 500\,$\mu$m thin AlN substrate end-cap. (ii) U-wire and Z-wire below the AlN substrate end-cap. (iii) Top view of a fiber-coupled side-cut nanowaveguide sample above the AlN substrate end-cap assembled with a gold deposited silicon substrate mirror (xy plane). (iv) Side view of U-MOT atoms (yz plane, no waveguide) (v) Side view of transported U-MOT atoms (yz plane, no waveguide); the U-MOT atoms are centered at $\sim$250\,$\mu$m from the surface, and the edge of U-MOT atoms is contiguous with the surface. The total number of U-MOT atoms ($N \simeq 5 \times 10^6$) decreases as the atom center moves closer to the surface. (c) A compact glass chamber with a fiber-coupled, integrated nanowaveguide atom-chip ($1.2 \times 1\,\mathrm{cm^2}$, see Fig.~\ref{fig_concept} (a)) for atom trapping, where nanowaveguides are fabricated along y-axis and MOT atoms are transported along z-axis.}
\label{fig_MOT_thermal_atom}
\end{figure}

By exposing the waveguide to a thermal vapor of Rb atoms, we were able to observe atomic absorption of the evanescent field of the waveguide (see Fig.~\ref{fig_MOT_thermal_atom} (a)), similar to that observed in a reference~\cite{Levy13}. The absorption spectrum is significantly broadened, both due to Doppler broadening and transit-time broadening through the sub-micrometer scale waveguide mode. To make this measurement we used a L-shaped nanowaveguide (shown in Fig.~\ref{fig_MOT_thermal_atom} (a) inset) to suppress light diverging through the $\mathrm{SiO_{2}}$ layer entering the output port, by more than 60\,dB.

We also produced the U-wire mirror MOT atoms (U-MOT) around the surface of a centimeter square gold mirror as shown in Fig.~\ref{fig_MOT_thermal_atom} (b), which is similar to an atom chip mirror MOT~\cite{Reichel99}. The U-MOT atoms are created by four cooling beams with the external U-wire (Fig.~\ref{fig_MOT_thermal_atom} (b)(ii)) and the Helmholtz bias coil; we chose the external U-wire instead of the microfabricated U-wire for simplicity. We transported the steady state U-MOT's center position toward the mirror surface by adjusting the magnetic field zero (see Fig.~\ref{fig_MOT_thermal_atom} (b)(v)), but found a U-MOT $\lesssim$50\,$\mu$m significantly decreased the atom number. Thus atom transport to the evanescent field atom trap a few hundred nm from the waveguide surface will be necessary. Microfabricated wires on the surface may be useful to bring atoms closer to the surface to allow loading efficiency into the evanescent field atom trap.

The detection of atomic absorption from the U-MOT was unsuccessful with the fiber-coupled nanowaveguide of the side-cut sample, as the shadow cast by residual silicon substrate nearby the nanowaveguide limited the cold atom density directly above the waveguide. This can be improved by using an integrated waveguide (Fig.~\ref{fig_MOT_thermal_atom} (c)), reducing the opening in the gold mirror to reduce the shadow, using some small angle beams to provide cooling at the waveguide surface, or dropping the atoms and switching on the waveguide field, or transferring the atoms to a purely magnetic trap.

In conclusion, we have designed a silicon nitride evanescent-field nanowaveguide platform for trapping Rb atoms with wavelengths of 760\,nm ($\lambda_{blue}$) and 1064\,nm ($\lambda_{red}$) and probing Rb atoms with the wavelength of 780\,nm ($\lambda_{\mathrm{Rb}}$). In a UHV environment, this nanowaveguide platform is compatible with the atom chip U-MOT configuration that will laser cool and confine atoms near the waveguide. The simulation of the nanowaveguide with inverse tapers shows high coupling efficiencies over a wide range of near-infrared wavelengths from 750\,nm to 1100\,nm. A propagation loss measurement based on a modified ``cut-back'' technique was used to confirm the high coupling efficiency per facet measured experimentally for the two-color atom-trap wavelengths. A centimeter-long nanowaveguide with a two-color evanescent field atom trap is expected to confine many atoms loaded from a mirror MOT, creating a high optical depth atomic sample with a strong atom-light interaction. As a step toward the realization of an evanescent field atom-light interaction, we measured an absorption profile of thermal atoms with a L-shaped nanowaveguide, and the absorption spectrum was strongly broadened due to Doppler broadening and transit-time broadening. In addition, we demonstrated U-MOT atoms in a compact UHV glass chamber, and the created U-MOT atoms are transportable to the surface. Such nanowaveguide atom traps may operate as an optically guided atom interferometer~\cite{Close13}, and can be used for inertial atomic sensors or atomic magnetometry. In general, the evanescent field nanowaveguide platform can also be used for bio-molecule sensing~\cite{Weisser99}, gas detection, and chemical solution sensing, which may be enhanced with dual color operation. Moreover, because of its good heat dissipation capability through its substrate, this scalable nanowaveguide platform has the potential for use in the implementation of collective-atom-based quantum networks.
%Ref.~\cite{Payne94}, Ref.~\cite{Lipson09}, Ref.~\cite{Zhou05},

\begin{acknowledgments}
This work is funded by ARO Atomtronics MURI project.
\end{acknowledgments}

%\nocite{*}
%\bibliography{aipsamp}% Produces the bibliography via BibTeX.

\begin{thebibliography}{99}
\bibitem{Porto03} J. V. Porto, S. Rolston, B. Laburthe-Tolra, C. J. Williams, and W. D. Phillips,``Quantum information with neutral atoms as qubits,''  Philos. T. Roy. Soc. A \textbf{361}, 1808, 1417–1427 (2003).
\bibitem{Lett09} A. M. Marino, R. C. Pooser, V. Boyer, and P. D. Lett, ``Tunable delay of Einstein–Podolsky–Rosen entanglement,'' Nature \textbf{457}, 859–862 (2009).
\bibitem{Tittle09} A. I. Lvovsky, B. C. Sanders, and W. Tittel, ``Optical quantum memory,'' Nat. Photon. \textbf{3}, 706–714 (2009).
\bibitem{Jessen99} G. K. Brennen, I. H. Deutsch, and P. S. Jessen, ``Quantum logic gates in optical lattices,'' Phys. Rev. Lett. \textbf{82}, 1060(1999).
\bibitem{Kimble08} H. J. Kimble, ``The quantum internet,'' Nature \textbf{453}, 1023 (2008).
\bibitem{Choi08} K. S. Choi, H. Deng, J. Laurat, and H. J. Kimble, ''Mapping photonic entanglement into and out of a quantum memory,'' Nature \textbf{452}, 67-71 (2008).
\bibitem{Kasevich13} S. M. Dickerson, J. M. Hogan, A. Sugarbaker, D. M. S. Johnson, and M. A. Kasevich, ``Multiaxis inertial sensing with long-time point source atom interferometry,'' Phys. Rev. Lett. \textbf{111}, 083001 (2013).
\bibitem{Biedermann11} H. J. McGuinness, A. V. Rakholia, and G. W. Biedermann, ``High data-rate atom interferometer for measuring acceleration,'' Appl. Phys. Lett. \textbf{100}, 011106 (2012).
\bibitem{Romalis03} I. K. Kominis, T. W. Kornack, J. C. Allred, and M. V. Romalis, ``A subfemtotesla multichannel atomic magnetometer,'' Nature \textbf{422}, 596–599 (2003).
\bibitem{Gaeta10} A. D. Slepkov, A. R. Bhagwat, V. Venkataraman, P. Londero, and A. L. Gaeta, ``Spectroscopy of Rb atoms in hollow-core fibers,'' Phys. Rev. A \textbf{81}, 053825 (2010).
\bibitem{Gaeta11} K. Saha, V. Venkataraman, P. Londero, and A. L. Gaeta, ``Enhanced two-photon absorption in a hollow-core photonic-band-gap fiber,'' Phys. Rev. A \textbf{83}, 033833 (2011).
\bibitem{Yang07} W. Yang, D. B. Conkey, B. Wu, D. Yin, A. R. Hawkins, and H. Schmidt, ``Atomic spectroscopy on a chip,'' Nat. Photon. \textbf{1}, 331–335 (2007).
\bibitem{Wu10} B. Wu, J. F. Hulbert, E. J. Lunt,	K. Hurd, A. R. Hawkins, and H. Schmidt, ``Slow light on a chip via atomic quantum state control,'' Nat. Photon. \textbf{4}, 776–779 (2010).
\bibitem{Spillane08} S. M. Spillane, G. S. Pati, K. Salit, M. Hall, P. Kumar, R. G. Beausoleil, and M. S. Shahriar, ``Observation of nonlinear optical interactions of ultralow levels of light in a tapered optical nanofiber embedded in a hot rubidium vapor,'' Phys. Rev. Lett. \textbf{100}, 233602 (2008).
\bibitem{Hendrickson10} S. Hendrickson, M. Lai, T. Pittman, and J. Franson, ``Observation of two-photon absorption at low power levels using tapered optical fibers in rubidium vapor,'' Phys. Rev. Lett. \textbf{105}, 173602 (2010).
\bibitem{Levy13} L. Stern, B. Desiatov, I. Goykhman, and U. Levy, ``Nanoscale light-matter interactions in atomic cladding waveguides," Nature Communications \textbf{4}, Article number: 1548 (2013).
\bibitem{Hinds11} M. Kohnen, M. Succo,	P. G. Petrov, R. A. Nyman, M. Trupke, and E. A. Hinds	
``An array of integrated atom–photon junctions,'' Nature Photonics  \textbf{5}, 35–38 (2011) doi:10.1038/nphoton.2010.255.
\bibitem{Lukin09} M. Bajcsy, S. Hofferberth, V. Balic, T. Peyronel,  M. Hafezi, A. S. Zibrov,  V. Vuletic, and  M. D. Lukin. ``Efficient all-optical switching using slow light within a hollow fiber,'' Phys. Rev. Lett. \textbf{102}, 203902 (2009).
\bibitem{Vetsch10} E. Vetsch, D. Reitz, G Sag{\'u}e, R. Schmidt, S. T. Dawkins, and  A. Rauschenbeutel, ``Optical interface created by laser-cooled atoms trapped in the evanescent field surrounding an optical nanofiber,'' Phys. Rev. Lett. \textbf{104}, 203603 (2010).
\bibitem{Lee15} J. Lee, J. A. Grover, J. E. Hoffman, L. A. Orozco, and S. L. Rolston, ``Inhomogeneous broadening of optical transitions of $^{87}$Rb atoms in an optical nanofiber trap,'' J. Phys. B: At. Mol. Opt. Phys. \textbf{48} 165004 (2015).
\bibitem{Lukin13} J. D. Thompson, T. G. Tiecke, N. P. de Leon, J. Feist, A. V. Akimov, M. Gullans, A. S. Zibrov, V. Vuleti{\'c}, and M. D. Lukin, {} ``Coupling a single trapped atom to a nanoscale optical cavity,'' Science \textbf{340}, no. 6137, pp. 1202-1205 (2013).
\bibitem{Kimble14} A. Goban, C.-L. Hung, S.-P. Yu, J. D. Hood, J. A. Muniz, J.H. Lee, M. J. Martin, A. C. McClung, K. S. Choi, D. E. Chang, O. Painter, and H. J. Kimble, ``Atom–light interactions in photonic crystals,'' Nat. Commun. \textbf{5}:3808 doi: 10.1038/ncomms4808 (2014).
\bibitem{Reichel99} J. Reichel, W. H{\"a}nsel, and T. W. H{\"a}nsch, {} ``Atomic micromanipulation with magnetic surface traps,'' Phys. Rev. Lett. \textbf{83}, 3398-3401 (1999).
\bibitem{Rolston13} J. Lee, D H. Park, S. Mittal, M. Dagenais and S. L. Rolston, ``Integrated optical dipole trap for cold neutral atoms with an optical waveguide coupler,'' New Journal of Physics. \textbf{15}, 043010 (2013).
\bibitem{Lipson03} V. R. Almeida, R. R. Panepucci, and M. Lipson,``Nanotaper for compact mode conversion,'' Opt. Lett. \textbf{28}, 1302 (2003).
%\bibitem{Payne94} F. P. Payne, and J. P. R. Lacey, {}``A theoretical analysis of scattering loss from planar optical waveguides,'' Optical and Quantum Electronics \textbf{26}, Issue 10, pp 977-986 (1994).
%\bibitem{Lipson09} A. Gondarenko, J. S. Levy, and M. Lipson, {} ``High confinement micron-scale silicon nitride high Q ring resonator,'' Optics Express \textbf{17}, Issue 14, pp. 11366-11370 (2009).
%\bibitem{Zhou05} K. Tsujikawa, K. Tajima, and J. Zhou, {} ``Intrinsic loss of optical fibers,'' Optical Fiber Technology \textbf{11}, Issue 4, Pages 319–331 (2005).
%\bibitem{Rauschenbeutel07} F. Warken, E. Vetsch, D. Meschede, M. Sokolowski, and A. Rauschenbeutel, {} ``Ultra-sensitive surface absorption spectroscopy using sub-wavelength diameter optical fibers,''  Optics Express \textbf{15}, Issue 19, pp. 11952-11958 (2007).
\bibitem{Hughes08} P. Siddons, C. S. Adams, C. Ge, and I. G. Hughes, {} ``Absolute absorption on rubidium D lines: comparison between theory and experiment,'' J. Phys. B: At. Mol. Opt. Phys. \textbf{41} 155004 (2008).
%\bibitem{Hasuo06} R. Kondo, S. Tojo, T. Fujimoto, and M. Hasuo, {} ``Shift and broadening in attenuated total reflection spectra of the hyperfine-structure-resolved D2 line of dense rubidium vapor,'' Phys. Rev. A \textbf{73}, 062504 (2006).
\bibitem{Close13} G. D. McDonald, H. Keal, P. A. Altin, J. E. Debs, S. Bennetts, C. C. N. Kuhn, K. S. Hardman, M. T. Johnsson, J. D. Close, and N. P. Robins, {} ``Optically guided linear Mach-Zehnder atom interferometer,'' Phys. Rev. A \textbf{87}, 013632 (2013).
\bibitem{Weisser99} M. Weisser, G. Tovara, S. Mittler-Nehera, W. Knolla, F. Brosinger, H. Freimuthb, M. Lacherb, and W. Ehrfeldb, ``Specific bio-recognition reactions observed with an integrated Mach–Zehnder interferometer, '' Biosensors and Bioelectronics \textbf{14}, Issue 4, pp. 405–411 (1999).
\end{thebibliography}

\end{document}